\documentclass[12pt]{iopart}
\usepackage{epsf}
\usepackage{amsfonts,amssymb}

\def\E{{\rm e}}
\def\I{{\rm i}}
\def\D{{\rm d}}

\begin{document}
\jl{1}

\title{Surprises in the suddenly-expanded infinite well}

\author{Claude Aslangul\footnote
  {{\bf e-mail:} aslangul@lptmc.jussieu.fr}}
\address{Laboratoire de Physique Th\'{e}orique de la Mati\`ere 
Condens\'ee, Laboratoire associ\'{e} au
CNRS (UMR 7600), \\ Universit\'e 
Paris 6, 2 place Jussieu, 75252 Paris Cedex 05, France}


%
\begin{abstract}

I study the time-evolution  of a particle prepared in the ground 
state of an infinite well after the latter is suddenly expanded. It 
turns out that the probability density $|\Psi(x,\,t)|^{2}$ shows 
up quite a surprising behaviour: for definite times, {\it plateaux}  
appear for which $|\Psi(x,\,t)|^{2}$ is constant on finite intervals 
for $x$. Elements of theoretical explanation 
are given by analyzing the singular component of the second 
derivative $\partial_{xx}\Psi(x,\,t)$. Analytical closed 
expressions are obtained for some specific times, which easily allow to 
show that, at these times, the density organizes itself into regular patterns 
provided the size of the box in large enough; more, above some 
critical time-dependent size, the density patterns  are independent 
of the expansion parameter. It is seen how the density at these  
times simply results from a 
construction game with definite rules acting on the pieces of the initial density. 
\end{abstract}


\submitted

\maketitle


\section{Introduction}\label{intro}

This short paper is devoted to some strange dynamical aspects around a problem 
which is often presented as the simplest one in quantum mechanics, 
namely the infinite one-dimensional well,  
although such a point of view can be seriously  questioned (for 
instance, what about the Heisenberg equations of motion for the 
infinite well?). Indeed, when going beyond academic elementary questions, 
this problem is {\it not} simple, and even turns out to be somewhat 
tricky, all 
subtleties obviously originating from the infinite discontinuities of the 
potential, which generates an infinity of bound states with an energy 
$E_{n}$ 
increasing without limit like the square of the quantum number $n$. 
This immediately entails that the propagator involves Gauss series 
(the Jacobi $\vartheta_{3}$-function being one very 
special case \cite{AbraSteg}), 
which are known to possess quite uncommon features; as an example, Holschneider \cite{Holschneider} 
shows that when the coefficients $c_{n}$ of the series are $\propto 
n^{-2}$, the sum is a self-similar function in a precisely defined 
sense. Here, I only aim to give a brief account of intriguing 
results, together with a far from being complete theoretical explanation 
and proof.  

To be sure and strictly speaking, infinite discontinuities can be  
discarded on physical grounds, 
but they conveniently mode\-lize a situation where the depth $V_{0}$ of the well is 
much greater than all other relevant energies, and where the space 
varia\-tion of the potential occurs on a length scale $l$  much smaller 
than  
all the others. Be it said in passing, for this reason, the classical limit of the 
infinite well is not a trivial point, due to the fact that one 
should first proper\-ly consider simultaneously the two limits $l\rightarrow 
0$ {\it and} $V_{0}\rightarrow +\infty$, in order to check whether they 
commute or not and, if they do not, to choose the physically relevant limiting 
procedure for the considered case (for another example, see 
\cite{appel}, \S\,1.6).

The problem at hand is the following.  Given that the particle (mass
$m$) is initially in an eigenstate of an infinite well, the well is
instantaneously expanded to a larger size: what is the subsequent
evolution of such a prepared initial state in the enlarged well?  Some
aspects of this problem have already been studied \cite{DoescherRice}
-- \cite{Pinder}; here I  focus on results which are absent of these
works and, up to my knowledge, seem unquoted in the literature. 
Obviously, any possible connection with an experiment would first of 
all require a 
proper analysis of various time scales, in order to be sure that the 
following 
theoretical framework is relevant to the experimental device.

Let us now enter into the specific problem and precise the notations 
used throughout. 
Taking, for the non-expanded well, $V(x)=0$ when $0<x<a$ and infinite elsewhere, the 
normalized eigenfunctions are:
\begin{equation}
    \psi_{n}(x)=\sqrt{\frac{2}{a}}\,\sin \frac{n\pi x}{a}\qquad(0\le 
    x\le a)\enspace,
    \label{eigensol}
\end{equation}
and vanish outside this interval; the eigenenergies are:
\begin{equation}
    E_{n}=\frac{n^{2}\pi^{2}\hbar^{2}}{2ma^{2}}\equiv n^{2}\hbar\omega_{1}
    \equiv n^{2}\frac{h}{T_{1}}\label{eigenenergies}\enspace.
\end{equation}
$n$ is a strictly positive integer, whereas $T_{1}$ is the smallest 
time-period of any time-dependent 
state built as a linear combination of the $\psi_{n}$'s.

From now on, I assume that, the particle 
being in the ground state $\psi_{1}(x)$ of the infinite well of width 
$a$,   
the latter is 
suddenly stretched at some time taken as $t=0$, 
increa\-sing in size from $a$ to $\lambda a$ with $\lambda >1$. Since the 
initial state $\Psi(x,\,0)\equiv\psi_{1}(x)$ is not a stationary 
state of the dilated well, $\Psi(x,\,t)$ has a non-trivial 
time-dependence and, among other things, expectation values of the 
obser\-vables which do not commute with the  Hamiltonian at $t>0$ show up actual 
time-dependence. I will focus on two of them, namely the probability 
density $\rho(x,\,t)$  and the density probability 
current $j(x,\,t)$ defined as usual:
\begin{equation}
    \rho(x,\,t)=|\Psi(x,\,t)|^{2}
    \enspace,\qquad     
    j(x,\,t)=\frac{\hbar}{m}\Im[\Psi^{*}(x,\,t)\partial_{x}\Psi(x,\,t)]\enspace,
    \label{denscourant}
\end{equation}
where $\Im$ denotes the imaginary part. $\rho$ and $j$ are related by the local conservation 
equa\-tion $\partial_{t}\rho+\partial_{x}j=0$. A few results 
concerning the averages of the position and the momentum of the 
particle will be briefly quoted at the end of the paper.

On the other hand, the expectation value of the 
energy does not change since no work is done on the particle when the 
well is expanded; this obvious physical fact 
will be analytically checked in due time. As for the variance of the energy, it 
vani\-shes before the expansion, but turns out to be infinite once the 
latter has been performed (see \mbox{section \ref{otherresults}}).

\section{Wavefunction at $t>0$
}\label{exppuitsinfp}

The eigensolutions of the expanded well are simply obtained by 
making $a\rightarrow\lambda a$ in formulas (\ref{eigensol}) and (\ref{eigenenergies}), namely:
\begin{equation}
    \psi_{\lambda,\,n}(x)=\sqrt{\frac{2}{\lambda a}}\,\sin \frac{n\pi 
    x}{\lambda     
    a}\equiv\frac{1}{\sqrt{\lambda}}\,\psi_{n}(\frac{x}{\lambda})\enspace,
    \label{eigensoldilate}
\end{equation}
for $0\le x\le \lambda a$, 
and:
\begin{equation}
    E_{\lambda,\,n}= \frac{1}{\lambda^{2}}E_{n}
    \label{eigenendilate}\enspace.
\end{equation}
Note that if $\lambda^{2}$ is an irrational number, the two spectra 
$E_{n}$ and $E_{\lambda,\,n}$ have no coincidence at all. The 
dilatation of the well lowers each eigenenergy, and yields an 
increased energy density (in infinite space, the spectrum is 
conti\-nuous). 

The resulting state at time $t>0$, $\Psi(x,\,t)$, can be developed on the complete 
eigenstates $\{\psi_{\lambda,\,n}\}_{n}$, and has an expansion of the form:
\begin{equation}    
    \Psi(x,\,t)=\sum_{n=1}^{+\infty}c_{n}\,
    \E^{\frac{1}{\I \hbar}E_{\lambda,\,n}t}\,\psi_{\lambda,\,n}(x)
    \label{psixt}\enspace.
\end{equation}
Note that, as thoroughly discussed by Styer\cite{Styer} in connec\-tion 
with the classical limit, it immediately results that the motion is periodic, with the 
period $T=\lambda^{2}T_{1}$, since the 
expansion of $\Psi(x,\,t)$ only contains integer multiples of the 
circular frequency $\omega_{\lambda}=\lambda^{-2}\omega_{1}$; as obvious on physical 
grounds, enlarging the well {\it increases} the period $T$ of the 
motion: for an infinite expansion, the motion is not periodic 
since, among other things, the wavepacket would spread out {\it ad 
infinitum}. Also note that the 
wavefunction at time $t$ is given by a Gauss series, {\it i.e.} a 
trigonometric series with time-oscillating factors of the form 
$\E^{\I  n^{2}\omega t}$, as contrasted to $\E^{\I  n\omega t}$ in a 
Fourier series. This yields quite rapid and irregular variations in time, all the 
more when the series coef\-ficients decrease slowly with $n$, which is 
the case here (see (\ref{coefinitexppl})).

As for the coefficients 
$c_{n}$, they are found by writing down the initial condition 
$\Psi(x,\,0)=\psi_{1}(x)$, and are thus equal to the scalar products 
$\langle\psi_{\lambda,\,n}|\psi_{1}\rangle$, namely:
\begin{equation}
    c_{n}=\frac{2}{a\sqrt{\lambda}}\int_{0}^{ a}\sin\frac{\pi
x}{a}\,\sin\frac{n\pi x}{\lambda a}\,\D x
    \label{coefinit}\enspace;
\end{equation}
note that the integral actually runs from $0$ to $a$, since $\psi_{1}(x)$ 
vanishes for any $x$ greater than $a$. A straightforward integration 
yields:
\begin{equation}
    c_{n}=\frac{2\lambda^{3/2}}
    {\pi}\frac{\sin\frac{n\pi}{\lambda}}{\lambda^{2}-n^{2}}
    \label{coefinitexppl}\enspace,
\end{equation}
so that the wavefunction at time $t\ge 0$ can be eventually written as:
\begin{equation}
    \Psi(x,\,t)=\frac{\I\lambda}{\pi}\sqrt{\frac{2}{a}}
\sum_{n=-\infty}^{+\infty}\frac{\sin\frac{n\pi}{\lambda}}{n^{2}-\lambda^{2}}\,
\E^{\I\frac{n\pi x}{\lambda a}}\,\E^{-\I n^{2}\omega_{\lambda} t}
\enspace,\label{Psiplusagr}
\end{equation}
for $0\le x\le\lambda a$, it being understood that $\Psi(x,\,t)$ vanishes 
outside the enlarged well.

For any given time $t$, $\Psi(x,\,t)$ is a continuous function of 
$x$ and of $t$; this is recognized from  
the fact that the coefficients $c_{n}$ behave like $n^{-2}$ for large 
$n$, ensuring that the series in (\ref{Psiplusagr}) is uniformly 
convergent. Obviously, this is not true for the $x$- or $t$-derivatives of 
$\Psi(x,\,t)$ (remember that the potential has {\it  infinite}  
discontinuities).

By construction, each exponential function $e_{n}(x,\,t)\equiv\E^{\I(\frac{n\pi x}{\lambda a}- 
n^{2}\omega_{\lambda} t)}$ satisfies the Schr\"{o}dinger equation 
$\I \hbar\partial_{t}e_{n}=-\frac{\hbar^{2}}{2m}\partial_{xx}e_{n}$, 
so that 
$e^{*}_{n}\partial_{xx}e_{n}-e_{n}\partial_{xx}e^{*}_{n}=0$: as 
it is the case for any stationary state in one dimension, the 
probability current constant in space, $\partial_{x}j_{\rm st}(x)=0$. 
This entails that performing a  
term-by-term derivation of the expansion (\ref{Psiplusagr}) to get 
the formal expression of the  current $j(x,\,t)$ related to 
$\Psi(x,\,t)$ can only generate singular terms, arising from the 
difference between the derivative of a function, and the series of the 
derivatives; these singularities turn out to be 
Dirac functions, which means that, for a given time, $j(x,\,t)$ is a 
piecewice constant  
function of $x$. Several examples of this will be given in due time.

Note that making $t=0$ in the RHS of (\ref{Psiplusagr}) leads to the function 
equal to $\sqrt{2/a}\,\sin(\pi x/a)$ for $0\le x\le a$, and equal to 
zero for $a\le x\le \lambda a$, since 
$\Psi(x,\,0)=\psi_{1}(x)$: in view of the sequel and considering the whole 
interval $[0,\,\lambda a]$, this allows to say (trivially at this 
point) that the initial 
probability density shows up a {\it plateau} with a vanishing value for $a\le x\le 
\lambda a$. From this, one concludes that the following equality holds 
true for any $x\in[0,\lambda a]$: 
\begin{equation}
    \frac{\I\lambda}{\pi}
\sum_{n=-\infty}^{+\infty}\frac{\sin\frac{n\pi}{\lambda}}{n^{2}-\lambda^{2}}\,
\E^{\I\frac{n\pi x}{\lambda a}}=\theta(a-x)
\sin\frac{\pi x}{a}
\enspace,\label{identities}
\end{equation}
where $\theta(x)$ is the unit step function ($\theta(x<0)=0$ and 
$\theta(x>0)=1$), as well as all the other equalities  obtained by a term-by-term
derivation; all of them can be $2\lambda a$-periodized in $x$ if 
needed. The important point to realize is that the series in the LHS of 
(\ref{identities}) is identically zero for any $x$ such that $a\le 
x\le \lambda a$. It turns out unnecessary to define the step function 
for $x=0$, since all the corresponding terms are multiplied by 
functions vanishing at this point.

As we will see, one strange thing (among others) is that the proba\-bility 
density at time $t$ 
also shows up {\it plateaux} (but not always with a {\it vanishing} 
value), in other finite intervals $[x_{k},\,x_{k+1}]$ at given 
periodic times; this can be figured out as the recurrent ghosts of 
the initial flatness on $[a,\,\lambda a]$.

Also note that if $\lambda\rightarrow 1$ (no change of the well), all 
coefficients go to zero, except for  $c_{1}$ which equals 1, as it must be. More generally, if $\lambda$ 
is a positive integer  $n_{0}$, the 
indetermination for $c_{n_{0}}$ is left by setting  $\lambda=n_{0}+\varepsilon$, and 
by taking the limit $\varepsilon\rightarrow 
0$. For $\lambda=n_{0}\in{\bf N}^{*}$, one thus obtains:
\begin{equation}
    c_{n_{0}}=\lim_{\varepsilon\rightarrow 
0}\,\frac{2(n_{0}+\varepsilon)^{3/2}}{\pi}\frac{\sin\frac{n_{0}\pi}{(n_{0}+
\varepsilon)}}{(n_{0}+\varepsilon)^{2}-n_{0}^{2}}\,=\,
\frac{1}{\sqrt{n_{0}}}\enspace.\label{limitcn0}
\end{equation}

Also note from (\ref{Psiplusagr}) that 
$\Psi(x,\,T-t)=\Psi^{*}(x, \,t)$, so that  
$\rho(x,\,t)=\rho(x,\,T-t)$: at times $t$ and  
$T-t$ the two density distributions coincide, but since the two wavefunctions 
are complex conjugate, the two correspon\-ding wavepackets have {\it opposite} group 
velocities; for the same reason the current satisfies 
$j(x,\,T-t)=-j(x,\,t)$. Other symmetry properties can be found by inspection of 
the series (\ref{Psiplusagr}); for example, one easily sees 
that 
for $t=T/4$, 
$\Psi(x,\,T/4)= -\Psi^{*}(\lambda a- x,\,T/4)$, namely that at 
a  quarter of the period (or at three-quarter), the density profile is even with regards to the middle  
of the dilated well. Other relations exist when both the abscissa and 
the time are changed, for instance one has 
\begin{equation}
\Psi(x,\, t+T/2)=-\Psi(\lambda 
a-x,\,t)
\label{symmtttx}
\end{equation}
for any $x$ and $t$. As we shall see, such symmetries play an 
important role, in particular to get closed convenient expressions for density and 
current at some remarkable times.

Since the initial state $\psi_{1}(x)$ is normalized to unity, so is $\Psi(x,\,t)$ at 
any time; this can be checked by a direct summation of the series 
$\sum_{n=1}^{+\infty}|c_{n}|^{2}$ (see Appendix A).

\section{Probability density plateaux and hints for a theoretical explanation
}\label{theoretexpla}

\begin{figure}[htbp]
\centerline{\epsfxsize=270pt\epsfbox{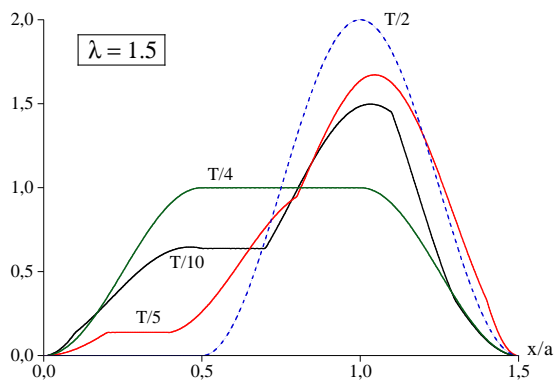}}
\vspace{-0pt}\caption{Probability density $|\Psi(x,\,t)|^{2}$ when the 
particle starts from the ground state of the undilated well; here, $\lambda = 1.5$. 
Each curve is labelled by the time $t$, $T$ being the 
period of the motion (see the text).}
\label{Psi2PuitsDil1.5}
\end{figure}

\begin{figure}[htbp]
\centerline{\epsfxsize=270pt\epsfbox{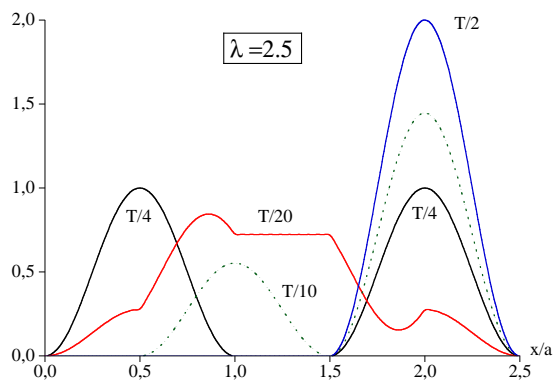}\hspace{15pt}}
\vspace{-0pt}\caption{Same as fig.\,\ref{Psi2PuitsDil1.5} 
for $\lambda = 2.5$.}
\label{Psi2PuitsDil2.5}
\end{figure}

\begin{figure}[htbp]
\centerline{\epsfxsize=220pt\epsfbox{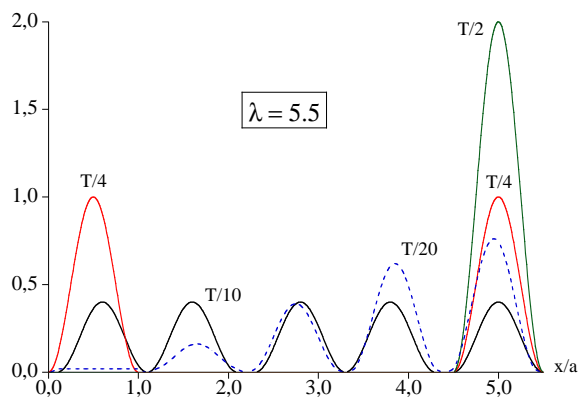}}
\vspace{-0pt}\caption{Same as fig.\,\ref{Psi2PuitsDil1.5} 
with $\lambda = 5.5$.}
\label{Psi2PuitsDil5.5}
\end{figure}

The surprise comes when plotting the probability density 
$\rho(x,\,t)$ at different times. Some examples are 
given in figs.\,\ref{Psi2PuitsDil1.5} -  
\ref{Psi2PuitsDil5.5}, which show that for very special times, 
the probability density assumes {\it constant values} in some definite 
intervals included in $[0,\,\lambda a]$. As already said, these {\it plateaux} can be 
figured out as the echoes of the flatness of $\Psi(x,\,0)$ with a 
zero height in the 
range $[a,\,\lambda a]$ for $x$.

 %
 %

The theoretical explanation of the existence of the {\it plateaux} 
lies on arguments which could be more firmly 
grounded if mathematical rigor were required. The basic idea is to use the second derivative 
$\partial_{xx}\Psi(x,\,t)$ as an indicator, 
since its singularities determine the absciss\ae\, where $\Psi(x,\,t)$ 
can have a cusp. Indeed, let us 
assume for definiteness that $x_{0}$ is 
an abscissa to the left of which $\Psi(x,\,t)$ is increasing, and 
is constant on the right. This means that the first derivative 
$\partial_{x}\Psi$ has a negative jump at $x_{0}$, entailing that 
the second derivative contains an additive singular term 
$\propto\delta(x-x_{0})$ with a negative weight, $\delta(x)$ being the 
Dirac function (remember that if a  
function $f(x)$ has a jump $\Delta f$ at $x=x_{0}$, its derivative is  
$f'(x)+\Delta f\delta(x-x_{0})\equiv f'(x)+D_{\rm sing}f$, where $f'$ 
is the ordinary derivative). At such a singular point, the 
derivative $\partial_{x}\Psi$ has a jump, so that, generally 
speaking, the density $|\Psi|^{2}$ shows up a cusp. Due to the general 
properties of the Schr\"{o}dinger equation, singularities are indeed 
to be expected in the second derivative in the presence of infinite 
discontinuities of the potential; they merely reflect, on a quantum 
mechanical level, the jumps of the veloci\-ty of a bouncing classical 
particle. These singularities, located at 
$x=0$ and $x=a$ at $t=0$ actually move 
about in the interval $[0,\,\lambda a]$ as time increases.

The second derivative of $\Psi(x,\,t)$ is obtained by a term-by-term 
derivation of the expansion  (\ref{Psiplusagr}); by writing 
$\frac{n^{2}}{n^{2}-\lambda^{2}}=1+ 
\frac{\lambda^{2}}{n^{2}-\lambda^{2}}$, it can be recast in 
the form:
\begin{equation}
         \frac{\partial ^{2}\Psi}{\partial x^{2}}=-\frac{\pi^{2}}{a^{2}}\Psi(x,\,t)+
        D_{\rm sing}^{2}\Psi
        \enspace,\label{psiseconde}
\end{equation}
where:
\begin{equation}
          D_{\rm sing}^{2}\Psi=-\frac{\I\pi}{\lambda 
a^{2}}\sqrt{\frac{2}{a}}\sum_{n=-\infty}^{+\infty}\sin\frac{n\pi}{\lambda}\,
\E^{\I n\frac{\pi x}{\lambda a}}\,\E^{-\I n^{2}\omega_{\lambda} t}
          \enspace\label{partsingder2}
\end{equation}
is the only singular part of $\frac{\partial ^{2}}{\partial
x^{2}}\Psi$.  The factor in the first term in the RHS of (\ref{psiseconde}) is 
recognized as $-\frac{2m}{\hbar^{2}}E_{1}$ and comes from the 
(ordinary) Laplacian 
operator in the (time-dependent) Schr\"{o}dinger equation. Now, having in mind the well-known Fourier expansion of the Dirac comb:
\begin{equation}
      \sum_{n=-\infty}^{+\infty}\E^{2\I\pi 
      nx}=\sum_{k=-\infty}^{+\infty}\delta(x-k)
      \enspace,\label{peignedeDirac}
\end{equation}
it is realized that $D_{\rm sing}^{2}\Psi$ embodies Dirac functions whenever 
the series in (\ref{partsingder2}) contains an infinite countable set of terms of the kind 
$\E^{\I\times{\rm integer}\times 2\pi}$, each having  a coefficient which is 
independent of the dummy summation label. In order to explore this 
possibility, I rewrite the expression (\ref{partsingder2}) as follows:
\begin{equation}
          D_{\rm sing}^{2}\Psi =\frac{\pi}{\sqrt{2}\,\lambda
a^{5/2}}\sum_{n=-\infty}^{+\infty}\left[\E^{\I\frac{n\pi}{\lambda
a}(x-a)}-\E^{\I\frac{n\pi}{\lambda
a}(x+a)}\right]\,\E^{-\I n^{2}\omega_{\lambda} t}
          \enspace.\hspace{5pt}\label{partsingder2bis}
\end{equation}
First of all, note that for $x=a$, the first series in 
(\ref{partsingder2bis}) reduces to $\sum_{n=-\infty}^{+\infty}\E^{-\I 
n^{2}\omega_{\lambda} t}$, {\it i.e.} generates a Dirac comb whenever $\E^{-\I 
n^{2}\omega_{\lambda} t}=1$ for an infinite countable set of values for $n$; this is the 
case if $t=(p/q)T$ with $p$ and $q$ integers: for all values of 
$n$ of the form $kq$ ($k$ integer), one has $n^{2}\omega_{\lambda} t=k^{2}qp\times 
2\pi$, which of the desired form: ${\rm integer}\times 2\pi$. 
At this stage, and considering only the first series in 
(\ref{partsingder2bis}), it is seen that  
a cusp {\it can} occur for $\Psi(x,\,t)$, with  
$(\partial_{x}\Psi)_{a+}-(\partial_{x}\Psi)_{a-}>0$ since the weight 
of $\delta(x-a)$ is then clearly a positive quantity.  Note that the same argument also holds for all 
the points of the form $\frac{x-a}{\lambda 
a}=$ even integer but all the corresponding absciss\ae \,are outside 
the relevant interval $[0,\,\lambda a]$ and may be ignored.

This tells us that $x=a$ is a good candidate, but this is just the beginning of the story, due to the existence of 
the second series in (\ref{partsingder2bis}). To show what can happen, 
let us set $x=a$ in both exponentials; the  whole series then writes 
\begin{equation}
     \sum_{n=-\infty}^{+\infty}\left(1-\E^{\I\frac{2n\pi}{\lambda
}}\right)\,\E^{-\I n^{2}\omega_{\lambda} t}
          \enspace.\label{partsingder2aa}
\end{equation}
In fact, it can happen that for all countably set of ``good'' 
values of the integer $n$, the two exponentials cancel each other, 
annihilating the possibility for the point $x=a$ to be a cusp. For definiteness, and as an example, let us go back to 
fig.\,\ref{Psi2PuitsDil1.5}, and consider the curve 
$t=T/2$ where the density clearly shows up a ``normal'' 
maximum at $x=a$. For 
this case, one has $p=1$ and $q=2$ in the above notations, which 
entails that the good values for $n$ are $n\,=2s$ 
($s$ integer); then, the only non-vanishing factors $(1-\E^{\I(2n\pi/\lambda)})$ are for $s=1,\,2\,(3)$, but for $s=1\,(3)$ 
and $s=2\,(3)$, they have 
opposite signs, so that the two related Dirac combs indeed have 
opposite weights, and the singularity at 
$x=a$ disappears. Thus, 
the point $x=a$ is not {\it always} such a remarkable point.

Let us now show that other values of the couple $(x,\,t)$ can define 
the edges of the {\it plateaux}, without trying to give an exhaustive 
catalogue of all these possibilities, just aiming at giving a few 
sufficient conditions for that.

The second exponential term $\E^{\I n\pi(x+a)/(\lambda a)}$ in (\ref{partsingder2bis}) is equal to $1$ for any $n$ if 
$(x+a)/(\lambda 
a)=r/s$, $r$ and $s$ integers, and if $n$ is an even multiple 
of  
$s$; the constraint $0\le x\le \lambda a$ entails $1\le 
r/s\le 1+1/\lambda$. This being realized, the conditions 
for the time-varying factor  $\E^{-\I 
n^{2}\omega_{\lambda} t}$ are the same as above, namely $t$ must be a rational 
fraction of the period $T$: 
$t=(p/q)T$. 
	
One example of such a case can be seen in fig.\,\ref{Psi2PuitsDil1.5}, where $\lambda=3/2$. For $t=T/4$ ($\omega_{\lambda} 
t=2\pi/4$), $p=1$, $q=4$ in the above notations. Close inspection 
reveals that $x=a$ is indeed a 
cusp, as well as  $x=a/2$ (take $r=s=1$); there 
is numerical evidence, and this is analytically proved below, that these 
points are in fact the edges of a {\it plateau}. 
Note that  the signs of the Dirac combs can be reversed; for instance 
with $(x+a)/(\lambda 
a)=r/s$, if $r$ is odd and  $n$ an {\it odd} multiple of  
$s$, $\E^{\I(2k+1)s\pi(r/s)}=\E^{\I(2k+1)r\pi}=-1$ (as  examples, see 
the curves $t=T/5$ and $T/10$ in fig.\,\ref{Psi2PuitsDil1.5}, 
for which the density increases on the left and to the right of the {\it 
plateau}).

Obviously, the existence of cusps is just a necessary condition for 
the occurrence of the {\it plateaux}. In order to analytically demonstrate 
their existence, one must generally prove that between two 
so identified given cusps, the density is indeed constant. This seems to be a rather 
intricate and difficult mathematical problem; in this short 
preliminary paper, I just intend to demonstrate this in a few 
specific cases, hoping to give a more complete and general proof in a 
future article.

\section{Closed expressions for specific times}\label{speciftimes}
It turns out that for some definite times $t_{k}$, closed expressions
of the wavefunction $\Psi(x,\,t_{k})$ can be written down.  I will here consider only
the three cases $t=T/2^{N+1}$ with $N=0,\,1,\,2$, before showing the
existence of quite another strange phenomenon, namely the
fragmentation of the wavepacket and the existence of regular patterns
when $\lambda$ is above a characteristic threshold $\lambda_{\rm c}$, depending on the
specific time considered.  The basic idea is to play with the time
phase factors appearing in the expansion (\ref{Psiplusagr}), and to
express $\Psi(x,\,t_{k})$ as a linear combination of the known initial
wavefunction taken at different absciss\ae \,\,$x_{i}$.  The
generalisation for times of the form $(p/q)T$ ($p$ and $q$ integers, 
$p<q$) seems quite
feasible, although it promises to be somewhat cumbersome as long as a 
more elegant method is not available.

A first observation is the following; at half of a period ($t=T/2$), a simple glance at the series 
(\ref{Psiplusagr}) allows to establish the following equality:
\begin{equation}
    \Psi(x,\,T/2)=-\Psi(\lambda a-x,\,0)
    \enspace,\label{psiTs2}
\end{equation}
which is just the symmetry relation (\ref{symmtttx}) for $t=0$; 
now, since $\Psi(x,\,0)$ is known (this is $\psi_{1}(x)$, which 
identically vanishes between $x=a$ and $x=\lambda a$), the 
equality   (\ref{psiTs2})
just gives a closed simple expression for the wavefunction at 
this remarkable time. In the following, I show how such a method can 
be used for the other times defined above.

\subsection{The case $t=T/4$}\label{persur4}

The case $\lambda=3/2$ (see fig.\,\ref{Psi2PuitsDil1.5}) and the 
spectacular {\it plateau} occurring for $t=T/4$ draws attention on 
this peculiar time. To start with and to introduce the method, let us 
analyze the things in details, but for any $\lambda$. The clue is 
simply to realize that for 
this peculiar time, the time-dependent exponential in (\ref{Psiplusagr}) is 
equal to 1 if $n$ is even, and \mbox{to $-\I $} if $n$ is odd. This 
allows to write $\Psi(x,\,T/4)$ in the form:
\begin{equation}
    \Psi(x,\,T/4)=S_{2,\,0}(x)-\I S_{2,\,1}(x)
    \enspace,\label{psidecT4}
\end{equation}
where the two (real) sums $S_{2,\,0}$ and $S_{2,\,1}$ respectively corres\-pond to even 
and odd values for the summation index $n$. Now that all the 
time-factors in the RHS of 
(\ref{psidecT4}) are fixed, it is tempting to compare such an 
expansion with $\Psi(x,\,0)$, which has a quite simple expression; observing that:
\begin{equation}
    \Psi(x,\,0)=S_{2,\,0}(x)+S_{2,\,1}(x)
    \enspace,\label{psidec000}
\end{equation}
and:
\begin{equation}
    \Psi(\lambda a-x,\,0)=-S_{2,\,0}(x)+S_{2,\,1}(x)
    \enspace,\label{psidec001}
\end{equation}
The two sums $S_{2,\,k}$ can now be expressed in terms of 
$\Psi(x,\,0)$, which is known, thus 
readily obtaining the sum of the series (\ref{Psiplusagr}) at this 
time:
\begin{eqnarray}
    \Psi(x,\,T/4)=\frac{1}{\sqrt{a}}[\E^{-\I\pi/4}
    \theta(a-x)\sin\frac{\pi x}{a}-\nonumber\\
   \hspace{140pt}\E^{+\I\pi/4}
    \theta(a-\lambda a+x)\sin\frac{\pi (\lambda a-x)}{a}]
    \enspace,\label{psiaT4}
\end{eqnarray}
an equality which yields the closed simple expression of 
the density for any $\lambda$:
\begin{equation}
    a\rho(x,\,T/4)=\theta(a-x)\,\sin^{2}\frac{\pi 
    x}{a}+\theta(a-\lambda a+x)\,\sin^{2}\frac{\pi (\lambda a-x)}{a}
    \enspace,\hspace{5pt}\label{Psi2LT4}
\end{equation}
with still $0\le x\le\lambda a$. 
Let me now choose $\lambda=3/2$; from (\ref{Psi2LT4}), it 
immediately results that:
\begin{equation}	
	a\rho(x,\,T/4)=\left\{ \begin{array}{lll}
	\sin^{2}\frac{\pi x}{a}\enspace, & \mbox{$ 0\le x\le a/2$} \\
	1\enspace, & \mbox{$a/2\le x\le a$} \\
	\cos^{2}\frac{\pi x}{a}\enspace, & \mbox{$a\le x\le 3a/2$}
	\end{array}\right.
	\label{psi2T4Lamb32}	
\end{equation}
which proves the existence of the {\it plateau} between $a/2$ and $a$ in 
this definite case. Note that this density is built in the following 
way: take the initial density, cut it into two pieces in the middle, 
translate the right part to the right of the distance $a$, draw a horizontal line 
between the two {\it maxima}, and divide the whole by a factor 2. We will 
recover such rules below, showing that the density at some other remarkable 
times can be built by playing with pieces of the initial density.

The expression (\ref{Psi2LT4}) is true at $t=T/4$ for any $\lambda$, 
and has two clearcut behaviours according to $\lambda<2$ or 
$\lambda>2$. In the first case, the two $\theta$ functions are 
simultaneously non-zero in the interval $[(\lambda-1)a,\,a]$, so that:  
\begin{equation}	
	a\rho(x,\,T/4)=\left\{ \begin{array}{lll}
	\sin^{2}\frac{\pi x}{a}\enspace, & \mbox{$ 0\le x\le (\lambda-1)a$} \\
	\sin^{2}\frac{\pi x}{a}+\sin^{2}\frac{\pi(\lambda a-  x)}{a}\enspace, & \mbox{$(\lambda-1)a\le x\le a$} \\
	\sin^{2}\frac{\pi(\lambda a-  x)}{a}\enspace, & \mbox{$a\le x\le \lambda$}
	\end{array}\right.
	\label{psi2T4Lambqcq}	
\end{equation}
This shows that for $1<\lambda<2$ but $\lambda\neq 3/2$, the function 
at $T/4$ has no {\it plateau}; in fact, for all such values of 
$\lambda$, numerical plots show that the latter does exist, but for other times 
and not located between the two simple values $a/2$ and $a$. 
Clearly the relative simplicity of the $\lambda=3/2$ case is due to 
the fact that $\lambda$ is a ``simple'' rational number. 

Note that the $x$-derivative 
of the density is equal to the real part 
$\Re(\Psi^{*}\partial_{x}\Psi)$; one easily checks from the 
expression (\ref{psiaT4}) that, for $\lambda<2$ (and still $t=T/4$),   
$\Re(\Psi^{*}\partial_{x}\Psi)$ never identically vanishes in a finite 
interval. Also note that if $\Psi(x,\,T/4)$ as given by (\ref{psiaT4}) is a continuous 
function of $x$ (as it must be), its  $x$-derivative is not, although it is devoid 
of Dirac peaks due to the cancellation of $\Psi(x,\,t)$ at each jump  
of the derivative: the density $\rho$, a 
continuous function of $x$, can indeed shows up cusps.

For $\lambda>2$, the two intervals $[0,\,a]$ and 
$[(\lambda-1)a,\,\lambda a]$ do not overlap; then, expression 
(\ref{Psi2LT4}) says that the wavefunction identically vanishes at 
$t=T/4$ for any $x\in[a,\,(\lambda -1)a]$. 
Examples of this are illustrated in figs.\,\ref{Psi2PuitsDil2.5} and \ref{Psi2PuitsDil5.5};  it is seen that $|\Psi|^{2}$ va\-nishes between $a$ 
and $3a/2$ for $\lambda=2.5$, between $a$ and $9a/2$ if $\lambda=5.5$. 
Thus, for $t=T/4$ and $\lambda>2$, the wavepacket splits in two 
distant parts: the particle is fully localized in two  
intervals separated by a finite one; in each interval, the profile is 
the clone 
of the initial one, just divided by 2. This {\it fragmentation} into identical 
curves will also 
occur for $t=T/8$: then, I will find four identical well-separated clusters, 
provided that $\lambda$ is greater \mbox{than $4$} (see section 
\ref{fragmentation}), each of them being one-quarter of the initial 
density $|\psi_{1}(x)|^{2}$ properly translated. 

\begin{figure}[htbp]
\centerline{\epsfxsize=280pt\epsfbox{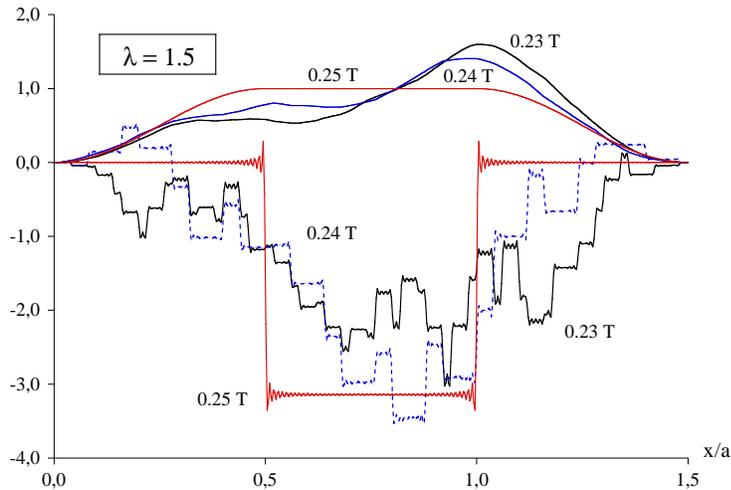}\hspace{15pt}}
\vspace{-0pt}\caption{Probability density (upper smooth curves) and 
density current (lower piecewise constant curves) for three very close 
times near $t=T/4$. Note the extreme variability of the current in space (except at 
exactly 
a quarter of a period). The 
oscillations near the jumps arise from numeri\-cal troncations of the 
series and are obviously related to some kind of Gibbs phenomenon 
adapted to a Gauss series.}
\label{Psi2Cour32}
\end{figure}

As explained above, the current probability density $j(x,\,t)$ is a
constant piecewise function, generally having jumps when the first
derivative of the wavefunction is discontinuous; otherwise stated, 
the jumps of $j(x,\,t)$ also occur whenever the singular part $D_{\rm 
sing}^{2}\Psi$ contains a Dirac comb. This turns out to happen in 
many points of the interval $[0,\,\lambda a]$, as seen in fig.\,\ref{Psi2Cour32}, where all the functions have been numerically 
computed from the series (\ref{Psiplusagr}). These plots  also show that there 
is not neces\-sary a 
direct relation between the jumps of the current and the edges of 
the {\it plateaux}, and reveals the irregular variation of 
$\partial_{x}\Psi$, which is not always clearly visible on the plot 
of the density, all the more since a cusp can occur only if 
$\Re(\Psi^{*}\partial_{x}\Psi)\neq 0$. 
Now, starting from 
(\ref{denscourant}) 
with $\Psi(x,\,T/4)$ given by (\ref{psiaT4}), a straightforward 
calculation yields the piecewice constant expression:
\begin{equation}
    j(x,\,T/4)=\frac{\pi\hbar}{ma^{2}}
    \,\theta(a-x)\theta[x-(\lambda-1)a]\,\sin\pi\lambda
     \enspace.\label{exprcouTs4}
\end{equation}
Again, the situation is quite different for $1<\lambda<2$ and for 
$\lambda >2$. In the first case, the current vanishes for 
$0<x<(\lambda-1)a$ and for $a<x<\lambda a$; in the intermediate 
interval, it assumes the constant negative value 
$\frac{\pi\hbar}{ma^{2}}\sin\pi\lambda$. Due to the conservation 
equation, the two points 
$x=(\lambda-1)a$ and $x=a$ are the only points where, at $t=T/4$, the 
time partial derivative $\partial _{t}\rho$ is non-zero. As 
contrasted, for $\lambda>2$, the current vanishes everywhere: not only 
at this time the wavepacket is split off in two fully disconnected parts, but the current 
between both regions is identically zero since there the wavefunction 
strictly vanishes.

\begin{figure}[htbp]
\centerline{\epsfxsize=300pt\epsfbox{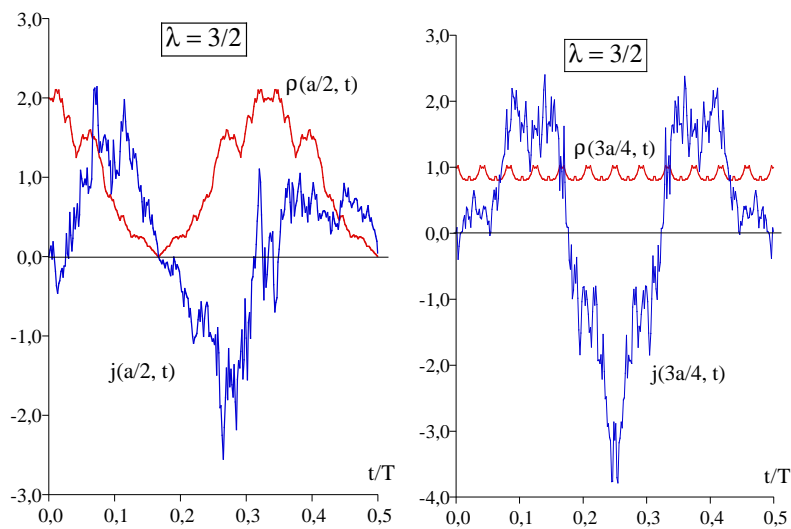}}
\vspace{-0pt}\caption{Probability density $\rho$ and current $j$ as a function 
of time at 
$x=a/2$ (left), middle of the well before the expansion, and $x=3a/4$ 
(right), middle of the well after the expansion.}
\label{Psi2CourT}
\end{figure}

I mentioned above that, due to the central role of the Gauss series 
given in (\ref{Psiplusagr}), it is expected that all 
quantities have a rather rapid and irregular variation in time. Such a fact is 
illustrated in fig.\,\ref{Psi2CourT}, where the proba\-bility density and 
current are plotted for a fixed $x$ as functions of time (remember 
that for $t$ and $T-t$ the densities are the same and the currents 
have reversed signs). At first 
glance, $j(x,\,t)$ even looks like a singular function; remember 
that $j$ is given by 
a double Gauss series. For 
$\lambda=3/2$, one has the symmetry $j(\frac{a}{2},\,t)=j(a,\,\frac{T}{2}-t)$.

\begin{figure}[htbp]
\centerline{\epsfxsize=300pt\epsfbox{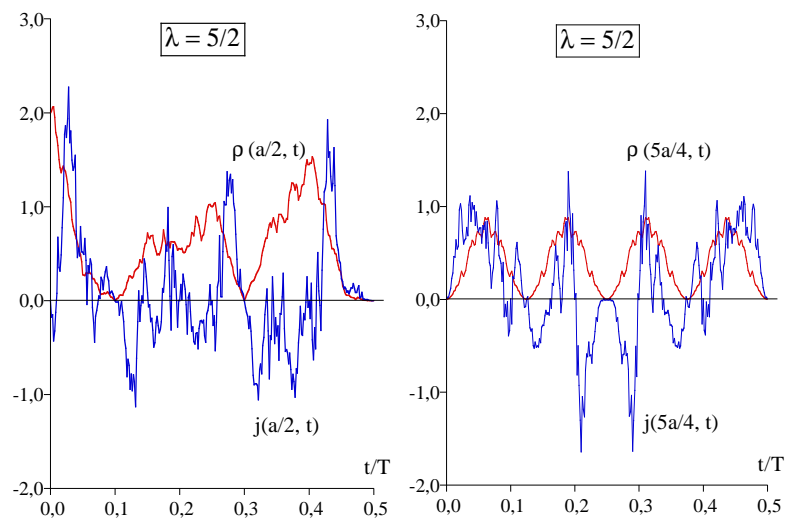}}
\vspace{-0pt}\caption{Same as fig.\,\ref{Psi2CourT} for $\lambda=5/2$.  
$x=a/2$ (left), middle of the well before the expansion, and $x=5a/4$ 
(right), middle of the well after the expansion.}
\label{Psi2CourT52}
\end{figure}

\subsection{The case $t=T/8$}\label{persur8}
I shall here follow the same arguments as before, the situation being 
a bit more complex. I first introduce four sums $S_{4,\,k}$ 
($k=0,\,1,\,2,\,4$) corresponding to the values $n=4p-k$ of the dummy 
summation 
variable in the series (\ref{Psiplusagr}). Now, inspection of the 
time phase factors shows that one has:
\begin{equation}
    \Psi(x,\,T/2^{N+1})=\sum_{k=0}^{2^{N}-1}\E^{-\I 
    k^{2}\pi/2^{N}}\,S_{2^{N},\, k}(x)
    \label{PsiTs2N}\enspace,
\end{equation}
which has now the form of a Gauss {\it sum}. For $N=2$, this gives:
\begin{equation}
    \Psi(x,\,T/8)=S_{4,\, 0}(x)-S_{4,\, 2}(x)+\E^{-\I\pi/4}[S_{4,\, 
    1}(x)+S_{4,\, 
    3}(x)]
    \label{PsiTs8debut}\enspace.
\end{equation}
I now follow the same idea as before, trying to choose definite 
absciss\ae\, $x_{i}$ such that the space factor in (\ref{Psiplusagr}) 
compensates in some way the dephasing due to the time factor. By 
trial and error, it is seen that the proper absciss\ae\, $x_{i}$ which allow to 
express 
the various sums in terms of the initial wavefunction 
$\Psi(x_{i},\,0)$ are 
$\lambda a/2\pm x$ and $3\lambda a/2-x$. First note that the sum  $S_{4,\, 1}(x)+S_{4,\, 
3}(x)$ is simply equal to the known quantity $S_{2,\, 1}(x)$ already introduced in 
subsection \ref{persur4}. As for the difference 
$S_{4,\, 0}(x)-S_{4,\, 2}(x)$, I find the following:
\begin{equation}
    \hspace{-2pt}0\le x\le\frac{\lambda a}{2}:\, S_{4,\, 0}(x)-S_{4,\, 
    2}(x)=\frac{1}{2}[\Psi(\frac{\lambda 
    a}{2}+x,\,0)-\Psi(\frac{\lambda 
    a}{2}-x,\,0)]
    \label{diffsommesxpetit}\,,
\end{equation}
\begin{equation}
    \hspace{-2pt}\frac{\lambda a}{2}\le x\le\lambda a:\, S_{4,\, 0}(x)-S_{4,\, 2}(x)=\frac{1}{2}[\Psi(-\frac{\lambda 
    a}{2}+x,\,0)-\Psi(\frac{3\lambda 
    a}{2}-x,\,0)]
    \label{diffsommesxgrand}\,.
\end{equation}
Great care must be exercized when writing the relations between the sums 
$S_{2^{N},\,k}$ and the values $\Psi(x_{i},\,0)$ due to the fact that the 
equality (\ref{identities}) {\it only} holds for $0\le x\le \lambda a$: outside this 
interval, the wave function vanishes, altough this is not the case 
for the sums since they are $2\lambda a$-periodic functions.

The above results eventually allow to  write the following closed expression for 
$\Psi(x,\,T/8)$ valid for any $\lambda$ (setting $\xi=x/a$ for 
simplicity):
\begin{eqnarray}
    \sqrt{2a}\Psi(x,\,T/8)=\theta(\frac{\lambda}{2}-\xi)f_{<}(\xi)+\theta(\xi-\frac{\lambda}{2})f_{>}(\xi)+\nonumber\hfill
    \\
    \hspace{70pt}\E^{-\I\pi/4}\left[\theta(1-\xi)\sin\pi\xi-
    \theta(1-\lambda+\xi)\sin\pi(\xi-\lambda)\right]
    \label{Psicoupee}\enspace,
\end{eqnarray}
where the two functions $f_{<}$ and $f_{>}$ are:
\begin{equation}
    f_{<}(\xi)=\theta(1-\frac{\lambda}{2}-\xi)\sin 
    \pi(\xi+\frac{\lambda}{2})+
    \theta(1-\frac{\lambda}{2}+\xi)
    \sin\pi(\xi-\frac{\lambda}{2})
    \label{finf}\enspace,
\end{equation}
\begin{equation}
    f_{>}(\xi)=\theta(1+\frac{\lambda}{2}-\xi)\sin\pi(\xi-\frac{\lambda}{2})+\theta(1-\frac{3\lambda}{2}+\xi)
    \sin\pi(\xi-\frac{3\lambda}{2})
    \label{fsup}\enspace.
\end{equation}
Note that small times give more cusps that larger times; numerical 
runs confirm that the initial two cusps propagate through the interval 
$[0,\,\lambda a]$ and multiply  at the very beginning of the motion, before 
reducing in number when the time gets closer to half of a period.

\begin{figure}[htbp]
\centerline{\epsfxsize=280pt\epsfbox{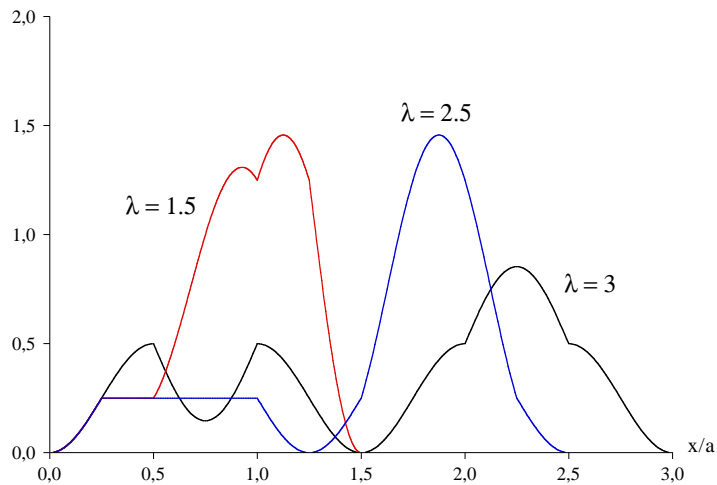}}
\vspace{-0pt}\caption{Probability density $\rho(x,\,T/8)$ 
calculated from the closed analytical expression (\ref{Psicoupee}), for 
three values of $\lambda$. Note the coincidence of the three densities for 
$0\le x\le a/4$, and between  $a/4$ and $a/2$ when $\lambda=1.5$ and $2.5$. 
The fact that the density is constructed with pieces of 
$|\Psi(x,\,0)|^{2}$ is clearly visible for the case $\lambda=3$.}
\label{rhoanalTs832253}
\end{figure}

In order to illustrate these results valid for any $\lambda$, let me 
again take $\lambda=3/2$; then, the above formula give for 
$2\sqrt{a}\Psi(x,\,T/8)$:
\begin{equation}
    0\le x\le\frac{ a}{4}:\, -(1+\I)\sin\frac{\pi x}{a}
    \label{Psi32a}\enspace,
\end{equation}
\begin{equation}
    \frac{ a}{4}\le x\le\frac{ a}{2}:\, -\I\sin\frac{\pi x}{a}-\cos\frac{\pi x}{a}
    \label{Psi32b}\enspace,
\end{equation}
\begin{equation}
    \frac{ a}{2}\le x\le a:\, -\I\sin\frac{\pi x}{a}-(2-\I)\cos\frac{\pi x}{a}
    \label{Psi32c}\enspace,
\end{equation}
\begin{equation}
    a\le x\le\frac{ 5a}{4}:\, -\sin\frac{\pi x}{a}-(2-\I)\cos\frac{\pi x}{a}
    \label{Psi32d}\enspace,
\end{equation}
\begin{equation}
    \frac{ 5a}{4}\le x\le\frac{ 3a}{2}:\, -(3-\I)\cos\frac{\pi x}{a}
    \label{Psi32e}\enspace.
\end{equation}
This  respectively gives the expressions for the dimensionless density 
$a\rho(x,\,T/8)$ in the corresponding five intervals:
\begin{eqnarray}
    \frac{1}2\sin^{2}\pi \xi\enspace,\quad\frac{1}{4}\enspace,\quad
    \frac{1}{4}+\cos^{2}\pi\xi-\frac{1}{4}\sin 
    2\pi\xi\enspace,\quad\nonumber
    \\
    \frac{1}{4}+\cos^{2}\pi\xi+\frac{1}{2}\sin 2\pi\xi\enspace,\quad
    \frac{5}{2}\cos^{2}\pi \xi
    \label{densiteTs8}\enspace;
\end{eqnarray}
note the {\it plateau} 
for $a/4\le x\le a/2$, and the cusp at $x=a$, all features which are 
apparent  in fig.\,\ref{rhoanalTs832253} where is plotted the density 
$a\rho(x,\,T/8)$ using the 
preceding formula, and the analytical expression (\ref{Psicoupee}) for 
the other $\lambda$ values. I checked that they 
give the same density as that obtained by a numerical 
calculation using directly the \mbox{expansion (\ref{Psiplusagr})}.

\begin{figure}[htbp]
\centerline{\epsfxsize=250pt\epsfbox{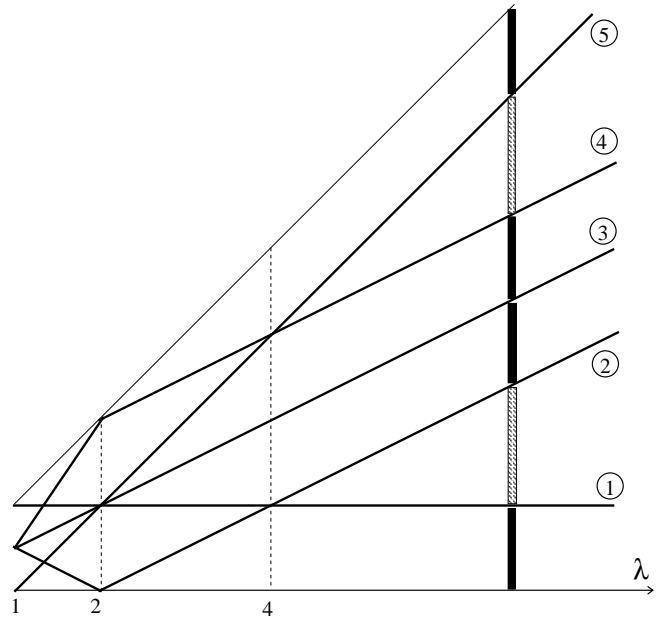}}
\vspace{-0pt}\caption{Absciss\ae\, of the cusps as a function of 
$\lambda$. The black segments show the domains where the density is 
non-zero; the hatched ones those where the density vanishes. 
Note that when $\lambda$ is above the threshold $\lambda_{\rm c}=4$, the domains of 
non-vanishing density move away one from the other, but keep the same 
size and shape.}
\label{AbscissesCusps}
\end{figure}

Coming back to the general $\lambda$ case, the expressions 
(\ref{Psicoupee}) - (\ref{fsup}) show that $\Psi(x,\,T/8)$ {\it a 
priori} shows up  cusps at the following absciss\ae,\, which I 
precisely define for further reference:
\begin{eqnarray}
   x_{1}=a\enspace,\quad x_{2}=|\lambda/2-1|a\enspace,
   x_{3}=\lambda a/2\enspace,\nonumber
   \\
   x_{4}=\theta(2-\lambda)(3\lambda 
   /2-1)a+\theta(\lambda-2)(1+\lambda/2)a\enspace,x_{5}=(\lambda-1)a
\label{cuspsabscisses}\enspace.
\end{eqnarray}
Quite remarkably, they are equally spaced, being 
located at 
$pa/4$ ($p=1,\,2,\,3,\,4,\,5$) for 
$\lambda=3/2$; for $\lambda>2$, where $(\lambda-1)a$ and $\lambda a/2$ 
merge, the cusp at $(3\lambda a/2-1)$ gets out of the interval 
$[0,\,\lambda a]$, but the cusp at $(1+\lambda/2)a$ comes in so that 
there is still 5 cusps, which all remain 
in the latter interval for any $\lambda$ (see fig.\,\ref{AbscissesCusps}). I will come back to this in 
the following subsection.

\begin{figure}[htbp]
\centerline{\hspace{30pt}\epsfxsize=290pt\epsfbox{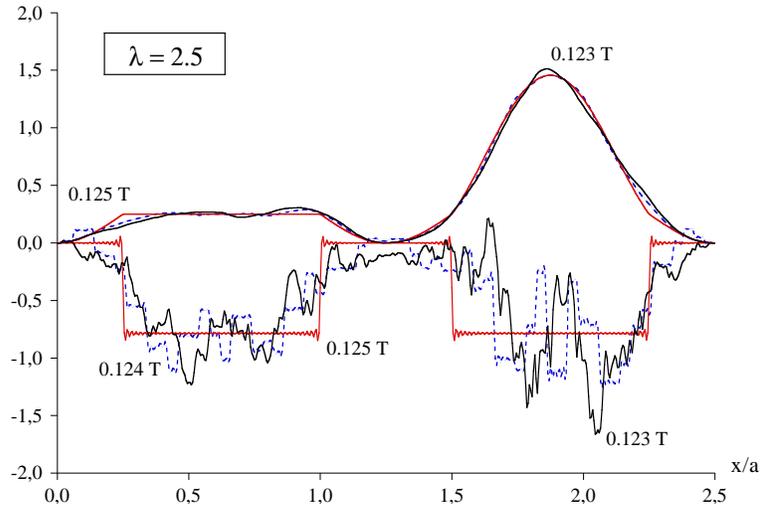}}
\caption{Probability density (upper smooth curves) and 
density current (lower piecewise constant curves) for three very close 
times near $t=T/8$).}
\label{Psi2Cour52}
\end{figure}

The current can also be easily computed; I find:
\begin{equation}    
    j(x,\,T/8)=\frac{\pi\hbar}{2\sqrt{2}\,ma^{2}}
    \left[c_{1\pm}(\xi)\sin\frac{\pi\lambda}{2}+c_{3\pm}(\xi)\sin\frac{3\pi\lambda}{2}\right]
    \label{courantTs8inf}\enspace,
\end{equation}
where the functions $c_{r\pm}$ depend on the considered interval; for 
$x<\lambda a/2$:
\begin{equation}
    \hspace{-2pt}c_{1-}(\xi)=\theta(1-\xi)[-\theta(1-\frac{\lambda}{2}-\xi)+\theta(1-\frac{\lambda}{2}+\xi)]
    +\theta(1-\frac{\lambda}{2}+\xi)\theta(1-\lambda+\xi)
\label{c1inf}
\end{equation}
and:
\begin{equation}
    c_{3-}(\xi)=\theta(1-\lambda+\xi)\theta(1-\frac{\lambda}{2}-\xi)
\label{c3inf}\enspace.
\end{equation}
For $x>\lambda a/2$, one has:
\begin{equation}
    \hspace{-2pt}c_{1+}(\xi)=\theta(1+\frac{\lambda}{2}-\xi)[\theta(1-\xi)+\theta(1-\lambda+\xi)]
    -\theta(1-\frac{3\lambda}{2}+\xi)\theta(1-\lambda+\xi)
\label{c1sup}
\end{equation}
and:
\begin{equation}
    c_{3+}(\xi)=\theta(1-\xi)\theta(1-\frac{3\lambda}{2}+\xi)
\label{c3sup}\enspace.
\end{equation}
All this shows that $j(x,\,T/8)$ is a piecewise 
constant function, as it must be. The density and the current are 
plotted in fig.\,\ref{Psi2Cour52} from the (truncated) series 
(\ref{Psiplusagr}) for three close times near $T/8$; note again the 
rapid variation of the current. For $\lambda>4$, the current 
vanishes everywhere.


\subsection{Fragmentation}\label{fragmentation}
One sees in figs.\,\ref{Psi2PuitsDil2.5} and \ref{Psi2PuitsDil5.5}, 
which both correspond to $\lambda>2$, that for $t=T/4$, the wavepacket 
is split into two symmetric parts at the edges of the allowed interval 
for $x$. This is true for any $\lambda>2$,  as a 
consequence of (\ref{Psi2LT4}): then, the two intervals $[0,\,a]$ and 
$[(\lambda-1)a,\,\lambda a]$ do not overlap, so that the density is 
non-zero only for $0<x<a$ and $(\lambda-1)a<x<a$; the two 
corresponding peaks are identical in shape, each equal to the initial density 
simply divided by $2$. It thus turns out that for times $T/4$ (and 
$3T/4$), the particle is fully localized into narrow domains and 
cannot be found between them. It can be said that, provided the 
expanded well has a size large enough, namely greater than $2a$, there 
is the possibility for two identical bumps of width $a$ localized at the 
edges of the box, with no density at all in between.

\begin{figure}[htbp]
\centerline{\epsfxsize=240pt\epsfbox{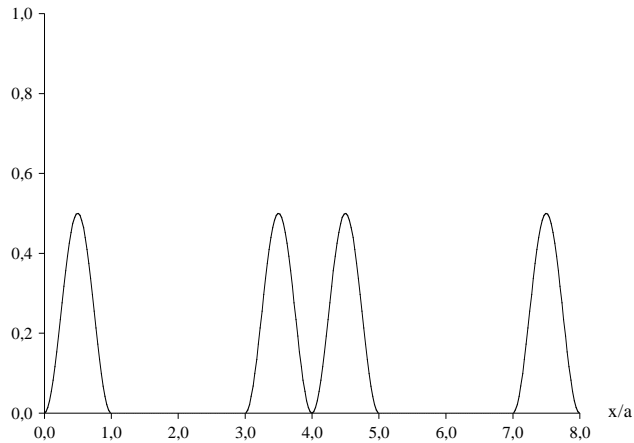}}
\vspace{-0pt}\caption{Probability density $\rho$ at $t=T/8$, for 
$\lambda=8$ above the critical value $\lambda_{\rm c}=4$. The fragmentation has 
occurred; the peaks now remain unchanged in size and shape when $\lambda$ varies 
and are located at the edges of the box, and on either side of the middle.}
\label{rhoanalTs8l8Gr}
\end{figure}

The same phenomenon occurs for $t=T/8$ (and $7T/8$): for $\lambda>4$, 
the density shows up four identical peaks, each of width $a$. Two of 
them are at the edges of the interval $[0,\,\lambda a]$, the two 
others are on each side of the middle of the box. Interestingly 
enough, the onset of the four peaks occurs at $\lambda=4$, a threshold 
at which the cusps are equally spaced (two couples of them are 
degenerate because here $\lambda-1=1+\lambda/2$ and $\lambda/2-1=1$). 
Once this has happened, the 
two middle peaks (``twin peaks'') remain at the fixed distance $a$ one 
from the other when 
$\lambda$ increases, being localized between $\lambda a/2\pm a$ 
(central cusps), while the 
two edge peaks also remain  unchanged and are still located between $0$ and 
$a$, and $(\lambda -1) a$ and $\lambda a$ 
as $\lambda$ varies (see fig.\,\ref{rhoanalTs8l8Gr}). It thus turns out 
that for $\lambda$ above the critical value $\lambda_{\rm c}=4$, the cusps delineate the regions of vanishing and 
non-vanishing density: $\lambda a/2$ and $(\lambda /2\pm 1)a$ for the 
central clusters, $a$ and $(\lambda -1)a$ for the ones localized near 
the boundaries of the box. Again, one can say that when the size is large enough 
(now greater than $4a$), four identical peaks of width $a$ can take place 
as indicated, and are independent of the 
expansion \mbox{parameter $\lambda$}.

To sum up this discussion, it can be stated that as far as $\lambda$ 
is greater than $4$, the density $\rho(x,\,T/8)$ is 
simply obtained 
by translating several times the initial density $|\psi_{1}(x)|^{2}\equiv\rho(x,\,0)$ 
according to the formula:
\begin{equation}
    \rho(x,\,T/8)=\frac{1}{4}\sum_{\alpha=1}^{4}\rho(x-l_{\alpha},\,0)\qquad 
    (\lambda>4)
    \label{profil}\enspace,
\end{equation}
where the location of the maxima $l_{\alpha}$ are $a/2$, $(\lambda\pm 
1) a/2$, and $(\lambda-1/2)a$. Increasing the expansion factor does 
not alters the profile of each peak; the twin peaks stay locked around the center of the 
box, whereas the edge peaks are getting more and more far away. 
Remember that above this threshold, the current identically vanishes. 

\begin{figure}[htbp]
\centerline{\epsfxsize=270pt\epsfbox{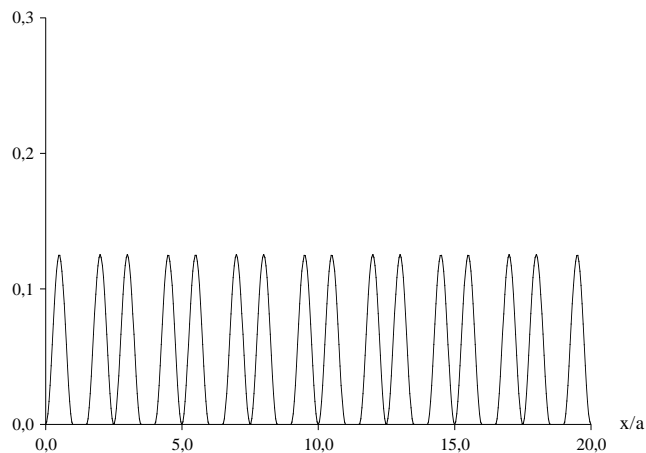}}
\vspace{-0pt}\caption{Probability density $\rho$ at $t=T/2^{N+1}$ 
avec $N=4$, for 
$\lambda=20$; here, the critical value is $\lambda_{\rm c}=2^{4}=16$.}
\label{rhoTS2N20PuitsDil}
\end{figure}

Gathering the above the results with those obtained in  the $T/4$ case, 
one can anticipate that for all times of the form $T/2^{N+1}$, there  
exists a threshold $\lambda_{\rm c}=2^{N}$ above which a fragmentation 
into $2^{N}$ peaks occurs. The density profile consists of the elementary pattern 
$\frac{1}{2^{N}}[|\Psi(x,\,0)|^{2}+|\Psi(\frac{\lambda 
a}{2^{N-1}}-x,\,0)|^{2}]$, and its $2^{N-1}-1$ clones translated by 
${\rm integer}\times\lambda a/2^{N-1}$; this 
is yet to be analytically proved in general, but numerical calculations 
allow to be convinced that this is true for any $N$ 
(see fig.\,\ref{rhoTS2N20PuitsDil} for an example). 
All this also confirms that many cusps exist at first times of the 
$T$-periodic motion, but 
remember that the time unit is precisely the period $T=\lambda^{2} T_{1}$, so that 
$t_{N}\equiv T/2^{N+1}=\lambda^{2}T_{1}/2^{N+1}\ge 2^{N-1}T_{1}$: 
large $N$ does not mean small times: clearly, the two functions  
$\Psi(x,\,T/2^{N+1})$ and 
 $\Psi(x,\,0)$ have no resemblance, although the latter allows to build 
the former according the above rules. 

The above conjectures are done in the continuity of the analytical 
results given in this paper. Many other statements can be claimed in 
view of numerical evidence, but they still remain to be proved; let me give a few of 
them:
\begin{enumerate}
    \item  For all times of the form $t_{M}=T/M$, $M$ integer, there exists a 
    threshold $\lambda_{\rm c}(M)$ above which complete fragmentation 
    occurs. 
    
If $M$ is even, $\lambda_{\rm c}(M)=M/2$ and one gets a 
    pattern of $M/2$ peaks located as above. If $M$ is odd, 
    fragmentation starts up at 
    $\lambda_{\rm c}=M$, with $M$ peaks; all peaks appear in twins except one, located 
    near the origin.

    \item  Fragmentation also takes place at times $pT/M$, with $p$ 
    integer. The number of peaks depends on whether $p$ and $M$ 
    have common divisors or not. For instance, with $M=12$, 
    $\lambda\ge \lambda_{\rm c}=6$, one finds six peaks if $p=1,\,5$, three peaks if 
    $p=2,\,4$, two peaks if $p=3$, and a single peak at $x=\lambda a$ 
    if $p=6$ (half-period).

\end{enumerate}
The  method presented in this paper should be 
still efficient for proving these (and other) statements, although 
a more elegant procedure is highly wishable in order to make the 
analysis less cumbersome 
and more systematic. Work in this direction is in progress.

\section{Other results}\label{otherresults}

After having focused on these rather outstanding behaviours, let me take the 
oppor\-tunity to add a few things for completeness, some of them being, 
as far as I know, unquoted in the literature.

\begin{figure}[htbp]
\centerline{\epsfxsize=200pt\epsfbox{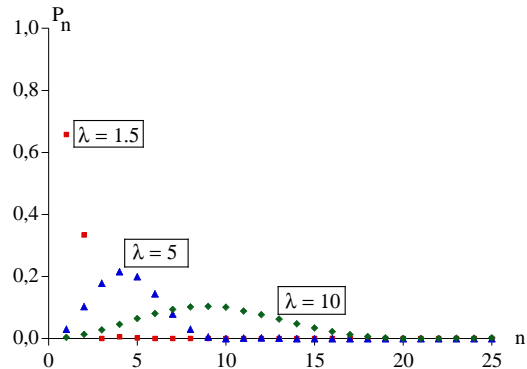}}
\vspace{-0pt}\caption{Probability distribution $P_{n}$ for 
three values of $\lambda$.}
\label{PnPuitsExpAJP}
\end{figure}

As a first by-product, one can compute the proba\-bility $P_{n}(t)$ to find the 
energy $E_{\lambda,\,n}$ when achieving a measurement of the energy at 
a time $t>0$; according to one of the postulates of quantum 
mechanics, one has 
$P_{n}(t)=|\langle\psi_{\lambda,\,n}|\Psi(t)\rangle|^{2}\equiv|c_{n}|^{2}$, namely:
\begin{equation}
    P_{n}=\frac{4\lambda^{3}}{\pi^{2}}\frac{1}{(\lambda^{2}-n^{2})^{2}}\,\sin^{2}\frac{n\pi}{\lambda}
    \label{probabn}\enspace;
\end{equation}
if $\lambda$ is equal to an integer $n_{0}$, the probability 
$P_{n_{0}}$ can be found from (\ref{limitcn0}), which 
yields $P_{n_{0}}=1/n_{0}$. 
When $\lambda\gtrsim 1$, the distribution of the $P_{n}$ is ever 
decreasing as a fonction of $n$; on the contrary, if $\lambda\gg 1$,
$P_{n}$ has maximum for $n\simeq \lambda$, but the probability 
distribution is quite flat (see fig.\,\ref{PnPuitsExpAJP}). This maximum 
has a clear meaning on 
physical grounds: there is some kind of resonance in the vicinity of the 
two states 
having an energy $E_{\lambda,\,n}$ close to $E_{1}$, the initial (and 
constant) value for the average energy. It can be checked that the 
expectation value 
$\sum_{n\in{\bf N}^{*}}P_{n}E_{\lambda,\,n}$ is indeed equal to $E_{1}$ 
at any time (see \mbox{Appendix A}).

Note that the variance of the energy is infinite, since the average 
$\langle H^{2}\rangle$ is given by a diverging series ($P_{n}\propto 
n^{-4}$, $E_{\lambda,\,n}^{2}\propto n^{4}$). This is due to the fact that the 
prepared state effectively implies a large number of eigenstates 
$\psi_{\lambda,\,n}$ because the coefficients $c_{n}$ have a 
slowly-decreasing algebraic  $n$-dependence, so that high energies are 
relevant. This yields divergent 
energy fluctuations. 

\begin{figure}[htbp]
\centerline{\epsfxsize=270pt\epsfbox{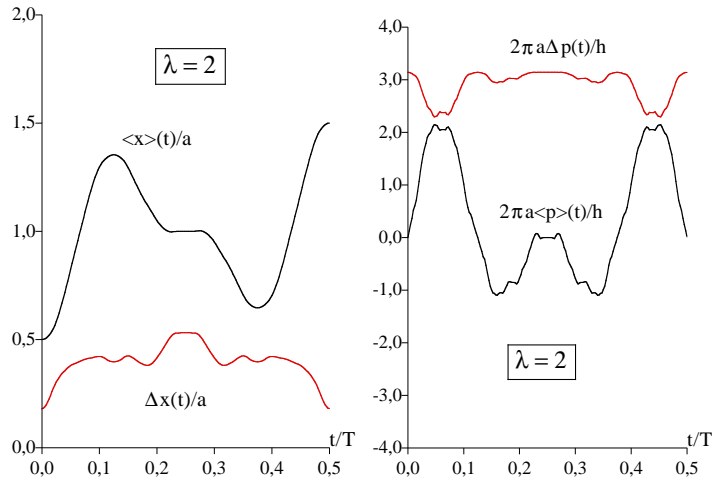}}
\vspace{-0pt}\caption{Left: variations in time ($0\le t\le T/2$) of the 
expectation value of the coordinate and of its variance. Right: same 
for the momentum.}
\label{X(T)P(T)2}
\end{figure}

\begin{figure}[htbp]
\centerline{\epsfxsize=240pt\epsfbox{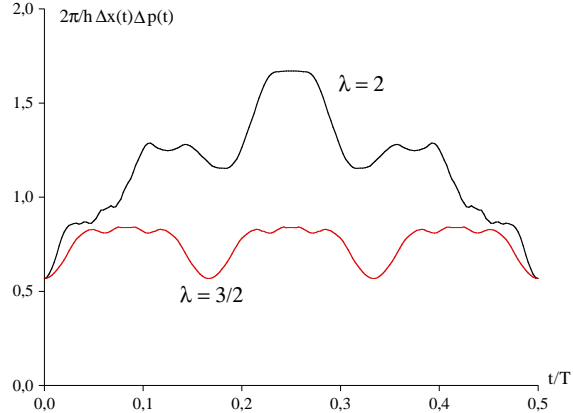}}
\vspace{-0pt}\caption{Variations in time ($0\le t\le T/2$) of the 
product $\Delta x\,\Delta p$.}
\label{DXDP}
\end{figure}

The expectation values of the position, $\langle x\rangle(t)$, and of 
the momentum, $\langle p\rangle(t)$, also display interesting behaviour 
with time. An example is given in fig.\,\ref{X(T)P(T)2}; it is seen 
that the particle  is periodically at rest on the average, since $\langle 
x\rangle(t)$ is constant and equal to $\lambda a/2$ whereas $\langle p\rangle(t)$ vanishes. This 
means that repeated measurements at those specific times would give 
exactly the same results as if the particle was in {\it any} 
stationary state of the dilated well. Measuring (independently) the 
energy would actually reveal the true nature of the state, giving for  
each measure one among all the possible energies $E_{\lambda,\,n}$. It 
is also numerically observed that $\langle x\rangle(t)$ is bounded 
by $a/2$ and $(\lambda-1/2)a$: the particle, in the average, gets 
never closer than $a/2$ to the reflecting walls at $x=0$ and $x=\lambda 
a$. The product $\Delta x\Delta p$ is plotted as a function of time 
in fig.\,\ref{DXDP}.

Note that the inverse process -- sudden compression of the 
well, $\lambda<1$ -- is impossible: one can not instantaneously generate 
a function vanishing for $\lambda a<x<a$ from a function which is 
finite in that interval. An infinite well can only be compressed with 
a {\it finite} rate; this case was analyzed in refs.  
\cite{DoescherRice} and\,\cite{Pinder}.

As a final remark, let me mention that the limit
$\lambda\rightarrow+\infty$ can be achieved from the above formula,
and indeed reproduces irreversible propagation in half-infinite space
starting from the initial state $\psi_{1}(x)$; the point is to observe
that $\Psi(x,\,t)$ in (\ref{Psiplusagr}) is a summation on the
variable $\nu=n/\lambda$, strictly equivalent to a Darboux sum, which 
quite 
naturally generates the Riemann integral over $\nu$ in this limit (the
differential element $\D\nu$ arises spontaneously from the factor
$1/\lambda$ in front of the summation).  From (\ref{Psiplusagr}) one
can thus write:
\begin{equation}
    \Psi(x,\,t)=\frac{\I\sqrt{2}}{a^{3/2}}\int_{-\infty}^{+\infty}
    \frac{\sin ka}{k^{2}-(\pi/a)^{2}}\,\E^{\I kx}\,
    \E^{-\I \hbar k^{2}t/(2m)}\D  k
    \label{psilambinf}\enspace.
\end{equation}
Note that the two zeroes of the denominator are just {\it apparent} 
singularities. Explicit direct calculation allows to check that such an 
expression coincides with that obtained directly with 
the propagator of a free particle in ${\bf R_{+}}$:
\begin{equation}
    U(x,\,t\,;\,x',\,0)=\frac{2}{\pi}\int_{0}^{+\infty}\sin 
    kx\sin k x'\,\E^{-\I\frac{\hbar k^{2}}{2m}t}\,\D k
    \quad(x,\,x'>0)\label{propagRsur2}
\end{equation}
acting on the initial state $\psi_{1}(x)$ to build the state at time
$t$ according to the standard way, namely $\Psi(x,\,t)=\int_{{\bf
R_{+}}}U(x,\,t\,;\,x',\,0)\Psi(x',\,0)\,\D x'$.  In Appendix B, I show
that the expression (\ref{psilambinf}) should not lead to
misconceptions about the $p$-representation of this wavepacket.

\section{Concluding remarks}
As stated from the beginning, this paper just aimed to present a brief 
review of the 
rather 
strange results given above. Although the general existence of the {\it 
plateaux} is numerically established, I was able up to this point to 
give only some elements of theoretical explanation, and a genuine proof 
in the two  
particular cases $t=T/4,\,T/8$. Clearly, further investigation is required in order 
to provide a general demonstration, and also to define a systematic method 
for finding the precise points $(x_{k},\,t_{k})$ in space-time where 
such intriguing behaviour takes place. 

The fragmentation phenomenon 
also requires more attention; at this point, it can be  
conjectured that for $t=T/2^{N+1}$, there exists a critical value 
$\lambda_{\rm c}=2^N$ above which spontaneous fragmentation occurs 
into $2^{N}$ peaks which are the translated {\it replica} of the 
initial density, divided by $1/2^{N}$; I gave an analytical proof only for 
$N=1,\,2$, but numerical evidence allows to be convinced that this is a 
general result. For $\lambda=\lambda_{\rm c}$, the density is an ordered  
finite lattice of adjacent bumps. It cannot be 
excluded that more complex patterns could be realized, going beyond 
the simple organization observed for $t=T/2^{N+1}$, although numerical 
calculations for times of the form $pT/M$ ($p$ and $M$ integers) have,  
until now, unveiled spatial organization having the simple features 
described above. Last but not least, a transparent 
physical interpretation would be welcome, allowing to get physical 
insight explaining such amazing and counterintuitive behaviours. Work in these directions is 
in progress and, hopefully, will be published in the near future.

\section*{Appendix A}

I here show how to check that the state $\Psi(x,\,t)$ given by
eq.(\ref{Psiplusagr}) is actually normalized to unity, and
that the expectation value of the energy is indeed equal to $E_{1}$ for
any time, as it must be on physical grounds since no work is done on
the particle when the well is suddenly expanded.

Let us consider the function $G(\lambda,\,\phi)$ defined as follows 
($\lambda$ not an integer):
\begin{equation}
     G(\lambda,\,\phi)=\sum_{n=-\infty}^{+\infty}
\frac{\E^{2\I  n\phi}}{\lambda^{2}-n^{2}}
     \enspace;\label{defGbetat}
\end{equation}
this series is uniformly convergent for any real $\phi$, so that $G(\phi)$
is a continuous function.  On the other hand, derivatives of $G$
obviously contain generalized functions (the unit-step function and
its derivatives).  One readily sees that the definition 
(\ref{defGbetat}) allows to write:
\begin{equation}
     |\langle\Psi(t)|\Psi(t)\rangle|^{2}=-\frac{\lambda^{2}}{2\pi^{2}}
     \left(\frac{\partial}{\partial \lambda}
     \left[G(\lambda,\,0)-G(\lambda,\,\phi )\right]\right)_{\phi=\pi/\lambda}
     \enspace.\label{normeGbetat}
\end{equation}

Let us now find $G(\lambda,\,\phi )$, which is an even $\pi$-periodic
function of the variable $\phi$.  By differentiating twice the definition
(\ref{defGbetat}), one obtains a linear combination of the function $G$ 
itself and of a
Dirac comb.  This means that the non-singular part of $G$ precisely
satisfies the differential equation $\partial_{\phi\phi}G+4\lambda^{2}G=0$
for any $\phi\in]0,\,\pi/2[$; the gene\-ral solution is $A\cos 2\lambda
\phi+B\sin 2\lambda \phi$.  The two constants $A$ and $B$ can be found by
using the known equalities (Mittag-L\ae ffler expansions):
\begin{equation}
     G(\lambda,\,0)\equiv\sum_{n=-\infty}^{+\infty}
\frac{1}{\lambda^{2}-n^{2}}=\frac{\pi}{\lambda}\cot\pi\lambda\enspace,
     \label{knownresidus}
\end{equation}
\begin{equation}
G(\lambda,\,\pi/2)\equiv\sum_{n=-\infty}^{+\infty}
\frac{(-1)^n}{\lambda^{2}-n^{2}}=\frac{\pi}{\lambda\sin\pi\lambda}\enspace,
     \label{knownresidus1}
\end{equation}
which yield $A=-\frac{\pi}{\lambda}\cot \lambda\pi$ and 
$B=\frac{\pi}{\lambda}$, so that eventual\-ly:
\begin{equation}
     G(\lambda,\,\phi )=\frac{\pi\cos 
     \lambda(2|\phi|-\pi)}{\lambda\sin\pi\lambda}\,\quad (-\pi/2\le \phi\le \pi/2)
     \enspace;\label{exprGbeat}
\end{equation}
As anticipated above, $G(\lambda,\,\phi )$ is a continuous function of $\phi$, but its first 
derivative has a jump at $\phi=0\,(\pi)$, explaining the presence of the Dirac comb in 
the complete second-order differential equation for $G(\lambda,\,\phi )$. 
Using now  
the rule expressed in (\ref{normeGbetat}), one readily gets 
$|\langle\Psi(t)|\Psi(t)\rangle|^{2}=1$. 

As for the average of the energy, 
one has:
\begin{eqnarray}
     \langle H\rangle=-\frac{\lambda E_{1}}{\pi^{2}}     
     \left[G(\lambda,\,0)-G(\lambda,\,\pi/\lambda)\right]-
     \hspace{60pt}\nonumber
     \\
     \hspace{120pt}
\frac{\lambda^{2}E_{1}}{2\pi^{2}}
     \left(\frac{\partial}{\partial \lambda}
     \left[G(\lambda,\,0)-G(\lambda,\,\phi )\right]\right)_{\phi=\pi/\lambda}
     \enspace;\label{moyEGbetat}
\end{eqnarray}
the quantity in the brackets of the first line vanishes since it is 
proportional to $\Psi(x=a,\,0)$; due to (\ref{normeGbetat}), one is eventually left with 
\begin{equation}
\langle H\rangle=E_{1}|\langle\Psi(t)|\Psi(t)\rangle|^{2}=E_{1}
\enspace,\label{relmoynorme}
\end{equation}
confirming that the
 expectation value of energy $\langle 
H\rangle$ is equal to $E_{1}$ at all times negative or positive, as it must be.

\section*{Appendix B}
I here intend to draw attention on a serious misconsception which could 
arise in view of the expression (\ref{psilambinf}). In order to make 
the discussion easier, I rewrite the latter as follows:
\begin{equation}
    \Psi(x,\,t)=\frac{1}{\sqrt{2\pi\hbar}}\int_{-\infty}^{+\infty}\E^{\I px/\hbar}
    \,{\tilde \Phi}(p)\,
    \E^{-\I p^{2}t/(2m\hbar)}\,\D  p
    \label{psilambinfbis}\enspace,
\end{equation}
where the function ${\tilde\Phi(p)}$ is:
\begin{equation}
    {\tilde \Phi}(p)=\frac{2p_{0}^{3/2}}{\I \pi}\frac{\sin(\pi p/p_{0})}{p_{0}^{2}-p^{2}}
    \label{philambinfaux}\enspace,
\end{equation}
where $p_{0}=\pi\hbar/a$. At first sight, it looks obvious to 
claim that ${\tilde \Phi}(p)$ is the $p$-representation of the 
initial state, while the time-dependent exponential in the integral 
in (\ref{psilambinfbis}) is just the 
ordinary phase factor for the free particle starting in the ${\tilde 
\Phi}(p)$ state at initial time. As apparently trivial as it stands, 
this statement is simply wrong. In order to show this, let us draw a few 
consequences of it.

First, it is easy to calculate the integral 
$\int_{-\infty}^{+\infty}|{\tilde \Phi}(p)|^{2}\,\D  p$; one finds 
that it is equal to $2$, instead of 1. Second, the true $p$-representation of the 
initial state can be easily and unambiguously calculated according to $\Phi(p,\,t=0)=(2\pi\hbar)^{-1/2}\int_{0}^{a}\E^{-{\rm 
i}px/\hbar}\,\sin(\pi 
x/a)\,\D  x$, and turns out to be:
\begin{equation}
    \Phi(p,\,0)=\frac{1}{\pi}\frac{p_{0}^{3/2}}{p_{0}^{2}-p^{2}}
    \left(1+\E^{-\I\pi p/p_{0}}\right)
    \label{philambinvrai}\enspace;
\end{equation}
aside the fact that it comes out properly normalized to unity 
since $\Psi(x,\,0)$ is, the function $\Phi(p,\,0)$ is frankly different from 
the function ${\tilde \Phi}(p)$ given in (\ref{philambinfaux}). 
Another drawback is that, due to standard rules of quantum mechanics 
for $p$-representation, the expectation value of the coordinate is:
\begin{equation}
    \langle x\rangle(t)=\I \hbar\int_{-\infty}^{+\infty}\D  p
    \,{\tilde \Phi}^{*}(p)[\frac{\D  }{\D  p}{\tilde \Phi}(p)-\frac{\I pt}{m\hbar}{\tilde 
    \Phi}(p)]
    \label{xmoyenfaux}\enspace.
\end{equation}
Since ${\tilde \Phi}(p)$ is an odd function of $p$, 
the integral vanishes, giving $\langle x\rangle(t)=0$, which 
is clearly incorrect: the wavepacket moves (and spreads out) in the free 
half-infinite space as time goes on. On the other hand, a 
non-vanishing integral would give a purely imaginaly expectation 
value since ${\tilde \Phi}(p)$ is a real-valued function, up to a 
constant phase.

The error comes from the fact that everything stands in ${\bf R}_{+}$, 
instead of ${\bf R}$. In other words, when a function $f(x)$ arises 
as a Fourier integral of the form:
\begin{equation}
    f(x)=\frac{1}{\sqrt{2\pi}}\int_{-\infty}^{+\infty}\E^{\I kx}
    \,{\tilde F}(k)\,\D  k
    \label{mifourier}\enspace,
\end{equation}
the equality holds true only for $x>0$ and one must not conclude at a 
glance (although this could happen to be correct) that the function ${\tilde F}(k)$ is the Fourier 
transform of $f(x)$: since all this holds true only if $x>0$, and assuming that 
the Jordan's lemma is applicable, one can add 
to ${\tilde F}(k)$ any function $\phi(k)$ which is analytic in the 
complex upper half-plane without changing the integral in the RHS of 
(\ref{mifourier}); the difference between ${\tilde \Phi}(p)$ and 
$\Phi(p,\,0)$ is actually such a function (remember that $\pm p_{0}$ 
are apparent singularities). In other words, although the Fourier transformation 
$f(x)\rightarrow F(k)$ is unambiguous, any intervening Fourier 
integral must be cautiously interpreted before to claim this is just the 
Fourier inversion 
formula; unconsidered intuitive 
identification can give incorrect results. Remind that for such functions 
defined in ${\bf R}_{+}$, the Laplace transformation is a much more 
secure method to proceed.

\ack

I am indebted to D. Mouhanna, J. Vidal, J.-M. Maillard, O. B\'{e}nichou  and R. Mosseri 
for helpful and fruitful discussions.

\end{document}